# Indirect Zero Field NMR Spectroscopy

**Authors:** Kai Buckenmaier[1,*], Richard Neumann[1], Friedemann Bullinger[1], Nicolas Kempf[1], Pavel Povolni[1], Jörn Engelmann[1], Judith Samlow[1], Jan-Bernd Hövener[2], Klaus Scheffler[1,3], Adam Ortmeier[4], Markus Plaumann[5], Rainer Körber[6], Thomas Theis[4] and Andrey N. Pravdivtsev[2]

*Corresponding author, Kai Buckenmaier kai.buckenmaier@tuebingen.mpg.de

**Affiliations:**

[1] High-Field Magnetic Resonance Center, Max Planck Institute for Biological Cybernetics; Tübingen, 72076, Germany

[2] Department Section Biomedical Imaging, Molecular Imaging North Competence Center (MOIN CC), Department of Radiology and Neuroradiology, University Medical Center Kiel, Kiel University, Am Botanischen Garten 14, 24118, Kiel, Germany

[3] Departement of Biomedical Magnetic Resonance, Eberhard-Karls University; Tuebingen, 72076, Germany

[4] Departement of Chemistry and Physics, NC State University; Raleigh, 27695, USA

[5] Institute for Molecular Biology and Medicinal Chemistry, Medical Faculty, Otto-von-Guericke-University; Magdeburg, 39120, Germany

[6] Physikalisch-Technische Bundesanstalt; Berlin, 10587, Germany

## Abstract

This study pioneers the two-field correlation spectroscopy (COSY) in zero to ultralow field (ZULF) liquid state NMR, employing hyperpolarized [1-$^{13}$C]pyruvate as a model system. We demonstrate the successful integration of signal amplification by reversible exchange (SABRE) for hyperpolarization, enabling the detection of ZULF COSY spectra with increased sensitivity. The use of field cycling allows the acquisition of two-field COSY spectra at varying magnetic field strengths, including zero-field conditions. This enables insight into both *J*-coupling and Zeeman-dominated regimes benefiting from ULF field observation sensitivity and mitigation of low-frequency noise by conducting readout at higher fields (>5 µT). Our study explores the effects of polarization transfer, apodization techniques, and the potential for further application of ZULF NMR in chemical analysis exemplified for three X-nuclei and three corresponding molecules: [1-$^{13}$C]pyruvate, [$^{15}$N]acetonitrile and [3-$^{19}$F]pyridine. These findings pave the way for more sensitive and cost-effective NMR spectroscopy in low-field regimes.

# Introduction

Nuclear magnetic resonance (NMR) is a powerful technique widely used in chemical analysis thanks to the high sensitivity of nuclear spins to their surroundings. Since its discovery in the 1970s, two-dimensional NMR spectroscopy(*1*) and the subsequent development of multi-dimensional NMR experiments(*2, 3*) have significantly advanced the field, improving NMR's ability to resolve complex spectra, leading to protein structure elucidation, and many other breakthroughs.(*4, 5*) Magnetic field cycling (MFC) is another technique that provides profound insights into molecular motion due to the dependence of relaxation parameters on the magnetic field.(*6, 7*) Inevitably, MFC and 2D NMR techniques were combined, leading to correlation spectroscopy (COSY)(*8*) and total correlation spectroscopy (TOCSY)(*9*) in two different magnetic fields. These techniques use different magnetic field strengths for signal evolution and detection e.g., benefiting from maximum sensitivity at high fields and functional spin properties at low fields. Due to hardware constraints of typical setups, the sample is physically moved between these different field strengths.

Zero and ultralow field (ZULF) NMR is gaining traction due to its spin-spin interaction regime and the capabilities that the low-frequency ZULF regime offers (**Figure 1a and b**),(*10–15*) such as NMR spectroscopy through metals,(*16*) Chemical analysis(*12, 17–20*) and ZULF-COSY.(*21*) However, these advancements are impossible with Faraday-based detection coils commonly used in high-field NMR due to their low sensitivity at low frequencies.(*22*) Instead, ZULF apparatuses (**Figure 1c**) use either superconducting quantum interference devices (SQUIDs)(*23, 24*) or optically pumped magnetometers (OPMs)(*25*) to detect the MR signal. OPMs have the advantage of not requiring cryogenics and directly observing MR signal. In contrast, SQUIDs have a much wider bandwidth, usually limited by the electronics driving the SQUID to the 1-10 MHz range for commercially available SQUID electronics and show also generally lower intrinsic noise levels than OPMs.(*26–28*) Crucially, both detectors measure magnetic flux independently of frequency, making them ideal for quantitative ZULF NMR investigations,(*23, 28*) in contrast to Faraday coils in high-field MRI, which measure the time derivative of the magnetic field. However, due to the intrinsic low polarization level of the sample at ZULF, usually, the sample also needs to be pre-(*29, 30*) or hyperpolarized.(*23, 31, 32*)

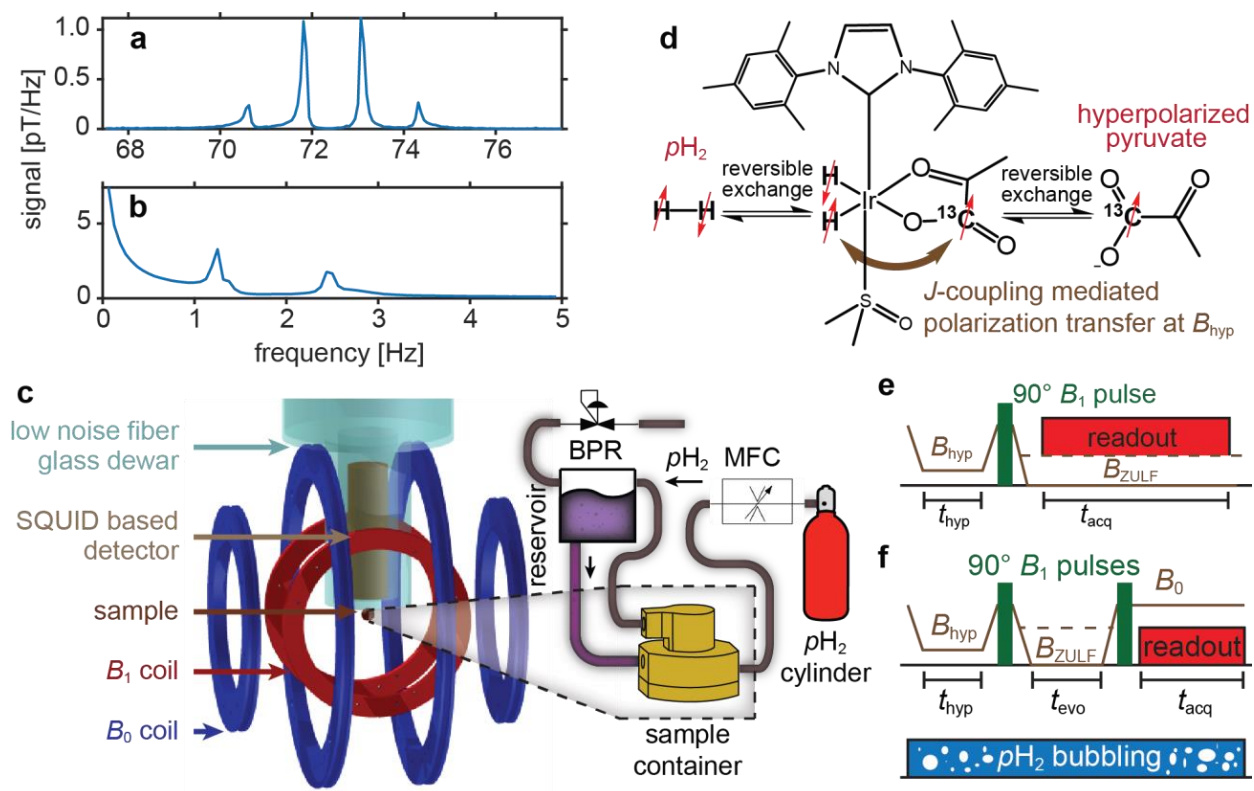

**Figure 1: ZULF experimental overview.** $^{13}C$ Phase-corrected NMR spectrum of [1-$^{13}C$]pyruvate at $B_{ZULF}$ = 6.8 µT (**a**) and $B_{ZULF}$ ≈ 0 nT (**b**). Experimental ZULF and SABRE-SHEATH setup (described in the methods section) together with the reactor section containing the $pH_2$ cylinder, the mass flow controller (MFC), sample container, the reservoir and the back-pressure regulator (BPR) (**c**). SABRE-chemistry schematic, where $pH_2$ and [1-$^{13}C$]pyruvate exchanges with an Iridium complex., At appropriate conditions (exposed to specific static magnetic fields or radiofrequency magnetic fields $B_{hyp}$) polarization transfer from $pH_2$ to $^{13}C$ occurs.(*33, 34*) For example, the target heteronuclei can be polarized, when the Larmor frequency difference between the $pH_2$ derived hydride ligands and the heteronuclei is of similar order to the combination of *J*-coupling interactions: this experiment is referred to as SABRE in shield enables alignment transfer to heteronuclei (SABRE-SHEATH).(*35*) (**d**). The ZULF FID sequence (**e**) and ZULF COSY sequence (**f**).

However, thanks to hyperpolarization techniques, conventional NMR is now accessible at lower target spin concentrations, reaching sub µM sensitivity at 14T (*36*) or sub mM sensitivity on benchtop spectrometers operating at 1 T.(*37*) Such incredible NMR sensitivity is achieved through enhancing the signal of nuclear spins by hyperpolarization, that temporarily increases the nuclear spin polarization of selected spins far beyond their thermal, Boltzmann-determined, equilibrium. The most commonly used hyperpolarization techniques include spin-exchange optical pumping (SEOP),(*38, 39*) dissolution dynamic nuclear polarization (dDNP),(*40*) Overhauser dynamic nuclear polarization (ODNP)(*41–43*), and parahydrogen ($pH_2$)-based hyperpolarization techniques including the hydrogenative variant of parahydrogen-induced polarization (PHIP),(*44*) and the non-hydrogenative variant signal amplification by reversible exchange (SABRE, **Figure 1d**).(*45–47*) PHIP and SABRE are relatively simple and inexpensive compared to commercially available dDNP with small-footprint cryogen-free polarizers(*48–50*) or polarizers placed within the bore of the MRI system.(*51*) In addition, SABRE enables stable hyperpolarization replenishment, which is ideal for 2D NMR.(*52*)

## Results

Using SQUID NMR and continuous SABRE in shield enables alignment transfer to heteronuclei (SABRE-SHEATH)(*35*) hyperpolarization, we indirectly measured the zero-field NMR spectra using the two-field MFC technique (**Figure 1f**). This approach takes advantage of the higher NMR sensitivity at higher frequencies since the noise level at ZULF is dominated by 1/f noise greatly increasing the noise at zero-field. The experiments with [1-$^{13}$C]pyruvate are detailed below, while experiments with [3-$^{19}$F]pyridine and [$^{15}$N]acetonitrile are presented in the supporting materials (**SI**). For signal observation, a home-built SQUID-based MFC NMR setup was used (**Figure 1c**, see description in methods).(*23*) Throughout all our experiments, pH$_2$ was continuously bubbled through the sample. In high-field NMR, such continuous bubbling would typically result in severe susceptibility artifacts; however, not at the ZULF regime.

In the following, we will introduce ZULF and the two-field MFC COSY NMR methods and then compare 1D and 2D COSY spectroscopy in a single and two fields. The experimental data are compared with corresponding simulations, as detailed in the methods section.

### 1D ZULF NMR spectroscopy

Before each readout phase, the sample was hyperpolarized using the SABRE-SHEATH technique. This was done by reducing the magnetic field for the time $t_{hyp}$ to the hyperpolarization field $B_{hyp}$. $B_{hyp}$ giving the highest signal of the X-nuclei was determined before for each substrate (see **Table S1** and **SI chapter 1**).

For the acquisition of 1D ZULF spectra (**Figure 1e, Table S2**), the magnetic field was first increased to $B_0 > 10$ µT where a 90° $B_1$ pulse in resonance with protons and X-nucleus was applied. After that, the field was reduced to the data acquisition field, $B_{ZULF}$, for the acquisition of free induction decay (FID) during $t_{acq}$ period. The signals Fourier transformation and phase correction lead to the 1D spectrum.

To anticipate the results from the two-field COSY experiment, we measured reference 1D NMR spectra. The FID recorded at $B_{ZULF}$ = 6.8 µT, in the Zeeman regime, i.e., when *J*-interaction between protons and X-nuclei is much smaller than the difference of their Larmor precession frequencies, shows the expected quadruplet with peaks separated by the *J*-coupling frequency of $^3J_{CH}$ = 1.25 Hz at the $^{13}$C Larmor frequency for [1-$^{13}$C]pyruvate (**Figure 1a**). In contrast, the FID recorded at $B_{ZULF}$ = 0 nT in the ZULF regime, i.e., when *J*-couplings exceed Zeeman interactions, results in a spectrum with peaks at 0, *J*, and 2*J*, all corresponding to the same *J*-coupling frequency of $^3J_{CH}$ = 1.25 Hz (**Figure 1b**). The zero-field spectrum was also used as a control to confirm the absence of residual magnetic field components that could affect the ZULF COSY spectra presented below. The spectra shown here agree with other ZULF spectra of [1-$^{13}$C]pyruvate found in the literature.(*12, 32, 53*)

### 2D ZULF NMR spectroscopy

A modified COSY readout scheme was used to acquire the ZULF COSY spectra (**Figure 1f**). Unlike conventional COSY sequences where the magnetic field remains constant at $B_0$,(*3, 52*) this modified approach lowers the $B_{evo}$ during the evolution period of length $t_{evo}$ between the two 90° $B_1$ pulses. For the readout phase, which lasted $t_{acq}$, the magnetic field was increased back to $B_0$, the same field strength when $B_1$ pulses were applied. $B_0$ was chosen to ensure that noise bands did not obscure any NMR signals. Consequent variation of the evolution time $t_{evo}$ (indirect dimension) and $t_{acq}$ (direct dimension), leads to 2D NMR spectra. By applying an apodization function, performing a 2D Fourier transform, and taking the absolute value (ref. (*54*), page 108), the 2D NMR spectrum is obtained (see methods and **SI chapter 3**).

In the following, we will compare two-field COSY spectra of [1-$^{13}$C]pyruvate measured at the *J*-coupling regime ($B_{evo}$ = 0 nT, *J*-coupling much larger than Zeeman splitting), intermediate coupling regime ($B_{evo}$ = 25 nT, *J*-coupling and Zeeman splitting are of the same order of magnitude) and Zeeman regime ($B_{evo}$ = 494 nT, Zeeman splitting much larger than *J*-coupling). The readout field was always the same and equal to $B_0$ = 6.8 µT, resulting in resonance frequencies of 72.4 Hz and 288.3 Hz for $^{13}$C and $^1$H respectively in the direct dimension.

## Two-field COSY of [1-$^{13}$C]pyruvate: *J*-coupling regime, $B_{evo}$ = 0 nT

The indirect dimension projection of the two-field COSY experiment with $B_{evo}$ = 0 nT (**Figure 2, Table S2**) results in a well-resolved spectrum with peaks at 0, ±*J*, and ±2*J* frequencies, as anticipated from the 1D ZULF NMR (**Figure 1b**). In the direct dimension projection, we observe the anticipated quadruplet peaks of [1-$^{13}$C], separated by the *J*-coupling frequency $^3J_{CH}$ = 1.25 Hz, as well as the expected doublet of $^1$H, also split by the *J*-coupling frequency. The peak positions in both the measured and simulated data show excellent agreement (**Figure 2**). By adjusting the simulation parameters (**Table S3**), the peak intensities were also well matched, and all features of the experimental data were successfully reproduced in the simulations.

It is important to note that in the measured $^1$H ZULF COSY spectrum, an additional peak appears at the symmetry center of the pattern (**Figure 2**, highlighted red box). This peak is due to orthohydrogen (*o*H$_2$), a by-product of the hyperpolarization reaction. (*55, 56*) It is clearly identified as *o*H$_2$ since no cross-peaks to pyruvate are observed. In the simulations, only [1-$^{13}$C]pyruvate was considered, which explains the absence of this peak in the simulated spectra.

Note that for measuring two fields of COSY within the *J*-coupling-dominated regime, it is essential that the residual field during the evolution period is below ≈ 2 nT. Even so, the whole SQUID-based NMR setup was situated within a three-layered shielding chamber (two layers of mu-metal shielding DC fields and one layer of aluminum for RF shielding), it was necessary to use additional 0th order shimming coils.

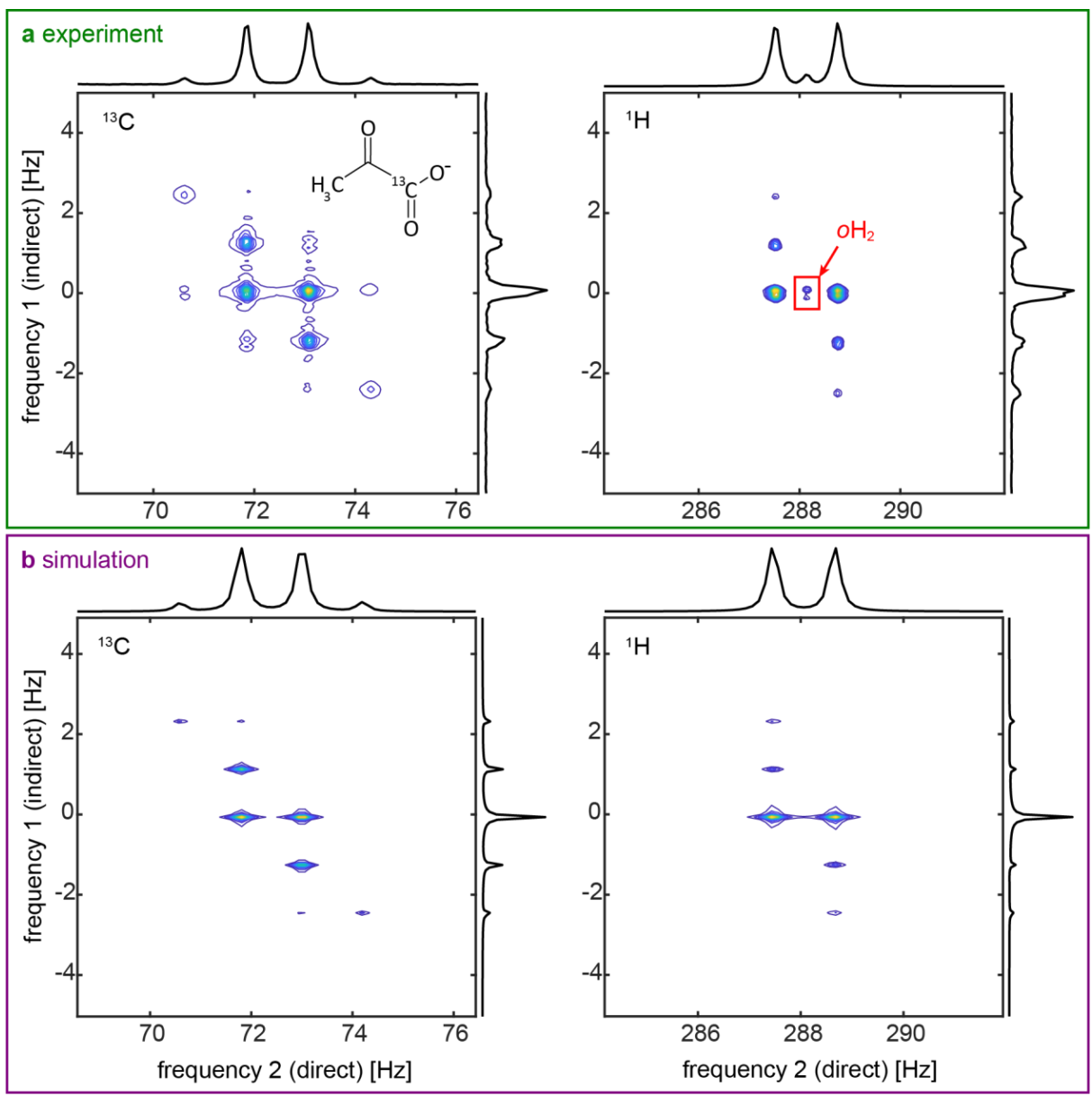

**Figure 2: Two-field COSY spectra of [1-$^{13}$C]pyruvate at $B_{evo}$ = 0 nT and $B_0$ = 6.8 µT.** Experimental (**a**) and simulated (**b**) absolute spectra of the $^{13}$C signal (left) and $^1$H signal (right) are shown together with projections in direct and indirect dimensions. We see peaks at the typical quadruplet positions of $^{13}$C and doublet positions of $^1$H. In the indirect frequency dimension projection, we see peaks at 0, ±$J$, and ±2$J$ frequencies.

### Two-field COSY of [1-$^{13}$C]pyruvate: Intermediate coupling regime, $B_{evo}$ = 25 nT

The evolution at an intermediate field of $B_{evo}$ = 25 nT results in a more complex spectrum because various spin transitions do not have matched resonance frequencies at high or zero field conditions. Note that no effect of the chemical shift is visible within this regime. Consequently, the peaks in the indirect dimension remain close to 0 Hz, but no distinct pattern is discernible in the 2D spectrum (**Figure 3a**). In the indirect dimension projection, the peaks are not clearly distinguishable, and the spectrum represents an evolution from the ZULF regime spectrum to the Zeeman regime spectrum. As with the ZULF COSY spectra at $B_{evo}$ = 0 nT, the peak positions of the experimental spectra at $B_{evo}$ = 25 nT show excellent agreement with the simulations.

In the direct dimension projection, the expected quadruplet peaks of $^{13}$C, separated by the J-coupling frequency $^3J_{CH}$ = 1.25 Hz, along with the doublet of $^1$H, also split by the J-coupling frequency, can be observed. For $B_{evo}$ = 25 nT, as expected, the $oH_2$ peak appeared at around 1 Hz (the proton Larmor frequency at $B_{evo}$) in the indirect dimension, placed between the proton's doublet in the direct dimension (**Figure 3a**, red boxes). An additional peak at 0 Hz is also observed in the indirect dimension, which is explained by the production of hyperpolarization during $t_{evo}$ at $B_{evo}$. This phenomenon is further described in the next section.

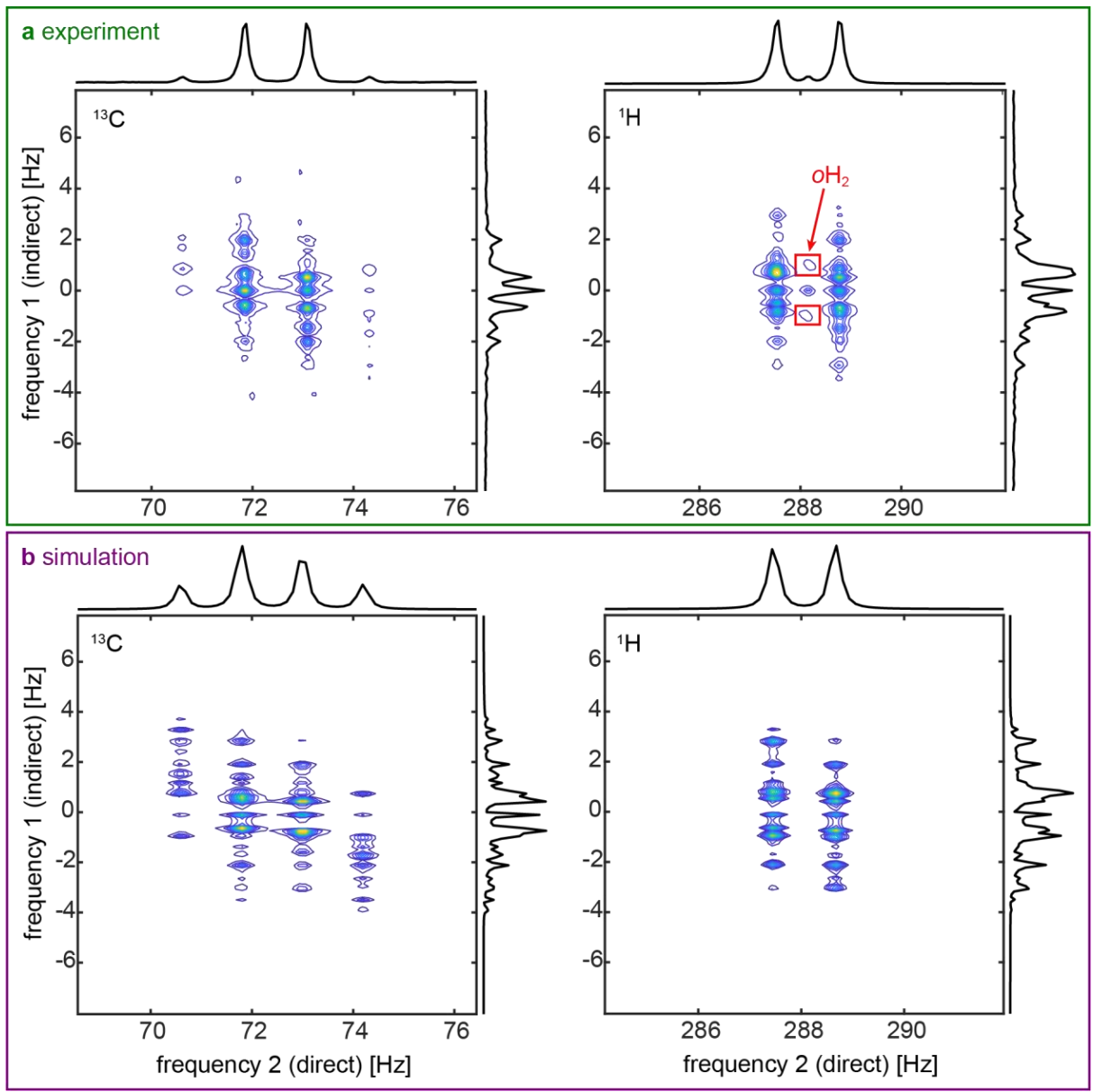

**Figure 3: Two-field COSY spectra of [1-$^{13}$C]pyruvate at $B_{evo}$ = 25 nT and $B_0$ = 6.8 µT.** Experimental **a** and simulated **b** absolute spectra of the $^{13}$C signal (left) and $^1$H signal (right) are shown together with projections in direct and indirect dimensions. We see peaks at the typical

quadruplet positions of $^{13}$C and doublet positions of $^1$H in the direct dimension. In the indirect frequency dimension projection, no clear pattern can be deduced in the intermediate coupling regime.

## Two-field COSY of [1-$^{13}$C]pyruvate: Zeeman regime, $B_{evo}$ = 494 nT

The magnetic field strength of the evolution field $B_{evo}$ = 494 nT falls within a range where the ZULF COSY spectrum displays Zeeman-dominated interactions. In this regime, the 2D spectrum exhibits symmetrical patterns, and the peaks can be attributed to direct and cross-peaks similar to a classical high-field heterogeneous COSY spectrum of [1-$^{13}$C]pyruvate (**Figure 4, Table S2**) with the difference that now both nuclei are measured simultaneously. Both projections of the spectrum on the direct and indirect frequency dimension reveal the anticipated quadruplet peaks at the $^{13}$C Larmor frequency and doublet structure for $^1$H with the *J*-coupling frequency $^3J_{CH}$ = 1.25 Hz.

Note that the $^1$H signal is significantly lower than that of $^{13}$C. As a result, the $^{13}$C-$^1$H cross-peak at the directly observed $^1$H signal exhibits greater intensity than the "diagonal" $^1$H peak. This discrepancy may arise from the fact that $^{13}$C is predominantly hyperpolarized via SABRE-SHEATH. A different pattern may emerge if conventional SABRE is employed for hyperpolarization, where the $^1$H pyruvate nuclei would be the primarily polarized nucleus.

An unexpected feature of this measurement is the presence of large peaks at 0 Hz in the indirect dimension of the experimental $^{13}$C signal (**Figure 4a**). During the evolution time between the two 90° pulses, $B_{evo}$ is set to 494 nT, which is close to the optimal hyperpolarization field $B_{hyp}$ of 350 nT. As a result, longitudinal polarization builds up during this period. It is subsequently excited by the second 90° pulse in the two-field COSY sequence, resulting in peaks located at 0 Hz in the indirect direction.

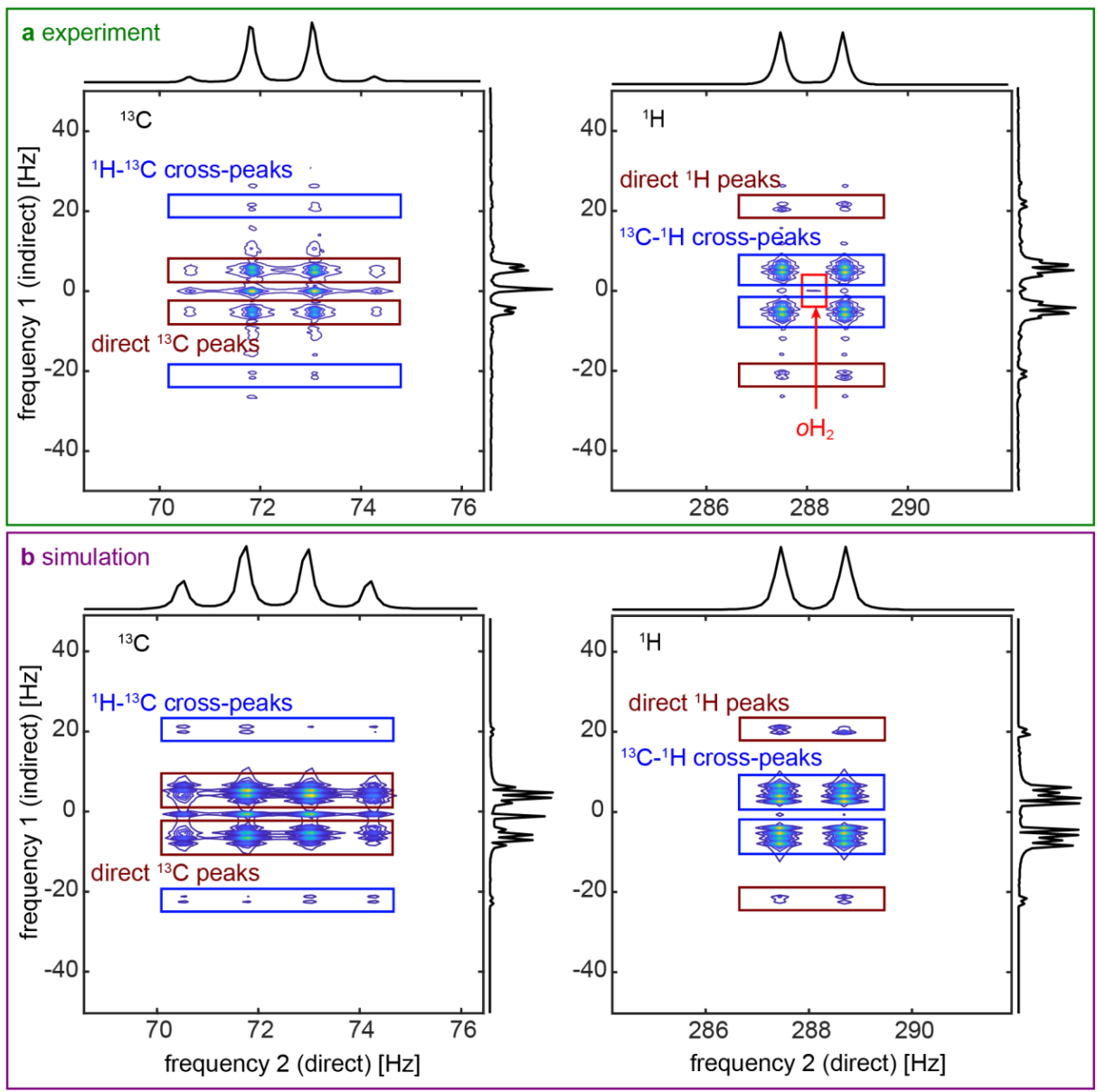

**Figure 4: Two-field COSY spectra of [1-$^{13}$C]pyruvate at $B_{evo}$= 494 nT and $B_0$ = 6.8 µT.** Experimental (**a**) and simulated (**b**) absolute spectra of the $^{13}$C signal (left) and $^{1}$H signal (right) are shown together with projections in direct and indirect dimensions. We see peaks at the typical quadruplet positions of $^{13}$C and doublet positions of $^{1}$H in the direct dimension. Indirect projections now comprise $^{1}$H doublet and $^{13}$C quadruplet, clearly assigning $^{1}$H-$^{13}$C and $^{13}$C-$^{1}$H cross-peaks. The order of nuclei in cross-peak assignment reflects the order of indirect and direct encoding.

# Discussion

In a previously published two-field NMR study containing a ZULF evolution field, the correlation between quadrupolar frequencies in a spin-1 system were measured for polycrystalline solids.(*57*) However, the frequency range of the spectra was at ≈ 100 kHz, in contrast to the 2D spectra demonstrated here with *J*-coupling signals at around 0 Hz.

Before conducting ZULF spectra, shimming must be employed as the shielding chamber has a low residual field. Measuring ZULF FID spectra by adjusting the shim coil currents leads to approximately 1 nT residual field. We used ZULF FIDs for shimming, and the ZULF COSY experiment took much longer to execute than conventional FID acquisition. So, it may not be reasonable to use ZULF COSY for ZULF FID acquisition if time consuming shimming is necessary.

The SNR of ZULF NMR systems is often limited by $1/f$ noise, with the most profound noise at zero frequency. This noise can originate from various sources, including the intrinsic noise of the magnetic field detector itself,(*22, 58*) noise from system components such as DC and AC power sources that drive the magnetic field coils, or unshielded external noise sources. The two-field COSY approach allows the acquisition of ZULF spectra at higher acquisition fields, offering the advantage of shifting the acquisition bandwidth to a region with lower intrinsic noise levels for NMR systems limited by $1/f$ noise at low frequencies. However, this ZULF-COSY approach comes with the trade-off of longer acquisition times compared to conventional NMR spectra (see **Table S2**).

It is important to note that for two-field COSY spectra with evolution time $t_{evo}$ close to zero-field conditions (< 100 nT), the bandwidth of the spectra is primarily determined by the *J*-coupling constant of the investigated substrate. This small bandwidth allows much faster acquisition of two-field COSY spectra compared to conventional COSY spectra. While the bandwidth of traditional COSY spectra can be reduced by folding higher frequencies into the bandwidth of the indirectly measured frequency dimension, the acquisition times remain significantly longer.(*52*)

Due to the low SNR at low frequencies without post-processing filters, as discussed above, evaluation was challenging. However, using the apodization function transforms difficult-to-analyze data into practically noiseless spectra presented above. More effort into the measurements and noise reduction would have been needed without the apodization function detailed in the **SI chapter 3.** At zero magnetic field, the spins do not precess on the plane like in high fields but instead oscillate between different spin states along a single axis.(*31, 32*) Therefore, it was critical to correctly calibrate the phase of the 90° pulse to acquire 1D ZULF spectra so that the axis of these one dimensional oscillations was parallel to the magnetic field detector (**Figures 1b and 1e**). Otherwise, the signal intensity would be significantly reduced. If the oscillation axis is perpendicular to the detector, no signal would be observed.(*31*) However, at the acquisition field $B_0$ of the two-field COSY sequence (**Figures 1e, 2–4**), the spin precession is sufficiently high that no phase adjustment of the 90° pulses is required.

In the intermediate regime of the ZULF-COSY experiment, the peaks of [1-$^{13}$C]pyruvate, which were located close to 0 Hz in the indirect dimension in the ZULF regime, drift outwards as the evolution field $B_{evo}$ is increased. Because now both the *J*-coupling and the Zeeman interaction are prominent in the indirect dimension, a complicated spectrum showing no distinguishable pattern is obtained.

In the Zeeman regime, the spectrum shows the expected conventional COSY characteristics. Due to having sufficiently high fields, the evolution field $B_{evo}$ and $B_0$, in both, direct and indirect, dimension, high field characteristics can be shown. The direct and cross-peaks can be clearly distinguished and assigned to the proton and $^{13}$C frequencies (**Figure 4**). This COSY behavior has already been shown multiple times(*59, 60*) but is a nice feature of the ZULF COSY sequence.

The fundamental/basic findings of above-discussed results were reproduced with another $A_3X$ system of [$^{15}$N]acetonitrile (**SI chapter 4.1, 4.2 and 5**). While in the ZULF regime, the spectrum pattern shows the same behavior as for the [1-$^{13}$C]pyruvate, only direct and cross-peaks of the $^{15}$N and not the proton are visible in the Zeeman regime. As the proton peaks in the ZULF COSY spectrum for the $^{13}$C already were low, for the [$^{15}$N]acetonitrile, the protons were insufficiently polarized to be observed. A ZULF COSY spectrum of [3-$^{19}$F]pyridine was also measured, but since the molecule is asymmetric with multiple different interactions, the spectrum is far too complex to discuss in much detail here (**SI chapter 4.3 and 5**).

Simulations can help to deduce the spin orders populated by hyperpolarization. Depending on the hyperpolarization procedure, different quantum coherences can be populated. For example, when pyruvate was hyperpolarized with dissolution dynamic nuclear polarization(*12*), the peak of the ZULF spectrum at *2J* had half the amplitude of the peak at *J*. This is different to what was observed for ZULF spectra of pyruvate hyperpolarized with SABRE-SHEATH (**Figure 1b** and ref. (*31*)). For SABRE, it strongly depends on the hyperpolarization conditions $B_{\mathrm{hyp}}$, $t_{\mathrm{hyp}}$ and methods (SABRE-SHEATH,(*61*) low field SABRE,(*62*) alt-SABRE,(*63*) LIGHT-SABRE,(*64, 65*) machete pulses(*66*)) to which degree the nucleus is hyperpolarized and to which degree the different quantum coherences are populated, resulting in different peak heights of the one- and two-dimensional NMR spectra. For the simulations, a combination of longitudinal single, two, and three spin states was considered to fit the measurements (**SI chapter 6**). Both the flip angle and polarization data, needed in the simulations to fit the experiment, stay identical for different ligands in the same regime but differ heavily in different regimes. The only difference in the experiment is the evolution field $B_{\mathrm{evo}}$. So, there's no experimental indication of why the flip angles vary significantly from the expected 90° pulse. We explained this effect by not considering hyperpolarization effects during the evolution phase. The production of hyperpolarization and relaxation at varying $B_{\mathrm{evo}}$ is different. This can be added explicitly by introducing a chemical exchange of $p$H$_2$ and substrate, as proposed previously(*67, 68*) but was not tested here.

## Conclusion

We were able to present two-dimensional two-field COSY for different evolution fields $B_{\mathrm{evo}}$, molecules and X-nuclei. An apodization method helped to significantly improve the quality of the acquired spectra. For $B_{\mathrm{evo}} \approx 0$ nT, we could show results where only the *J* coupling is the relevant variable responsible for the indirect dimensional spectrum. In combination with the high-field spectrum in the direct dimension, the two-dimensional figure showed an asymmetric peak pattern, revealing connections between different transition frequencies of the ZULF and Zeeman regimes. The result of the indirect dimension projection is the desired zero-field spectrum. Deviations of the expected spectrum can indicate the magnetic field inhomogeneity or deviation from the zero-field and can be used to measure previous. The method can become an alternative to the traditional direct measurement of zero-field NMR. Although we have used the SQUID system to measure the signal here, we envision that, using higher magnetic fields and simple Farraday coils, zero-field spectra can be measured using the two-field COSY method discussed, making ZULF spectroscopy more sensitive and accessible.

## Methods

### Setup and Reactor:

The experimental setup is housed within a three-layered shielding chamber, maintaining a residual magnetic field of approximately 10 nT. Consequently, magnetic field shimming was necessary to acquire ZULF COSY spectra and 1D ZULF NMR.

The zero and ultralow field setup (**Figure 1c**) consists of a SQUID-based magnetic flux detector coupled to a second-order gradiometer. The sensor sits within a low-noise fiberglass dewar under which the sample reactor is positioned with a hot-to-cold distance of about 13 mm. A tetra coil was used to generate the fields $B_{hyp}$, $B_{evo}$, $B_{ZULF}$, and $B_0$. The $B_1$ RF field was generated with a Helmholtz coil. Directly on the $B_1$ coil sits an additional shimming coil.

The sample reactor is a closed system comprising a sample container and a reservoir, allowing long-lasting experiments. The sample container holds the sample while $pH_2$ is bubbled through it. Additionally, the sample chamber is continuously replenished by the connected reservoir. The system can be pressurized up to 10 bars using a combination of a back-pressure regulator (BPR, P-7XX, IDEX series of back pressure), a mass flow controller (MFC), and pressurized $pH_2$. The setup without this pressure option is described in Ref. (*23*).

### $pH_2$ generator

$pH_2$ generators operate at cryogenic temperatures, where a paramagnetic catalyst, typically iron oxide, converts normal dihydrogen gas into $pH_2$-enriched gas. Normal dihydrogen consists of approximately 25% $pH_2$ and 75% $oH_2$. The degree of $pH_2$ enrichment is temperature-dependent, with enrichment levels ranging from 50% at 77K to over 99% below 25 K. Using a commercially available system (XEUS-technologies)(*69*), we achieved about 95% $pH_2$ fraction. We apply a cooling temperature of 25 K, slightly above the boiling point at ambient hydrogen pressure of 20 K. The $pH_2$ was stored in aluminum gas cylinders at a pressure of approximately 30 bar.

### Sample preparation

**Pyruvate sample:** The sample consisted of 50 mM sodium [1-$^{13}$C]pyruvate (Sigma-Aldrich, article no. 490709), 5 mM [Ir(COD)(IMes)Cl] precatalyst (COD = 1,5-cyclooctadiene, IMes = 1,3-bis(2,4,6-trimethylphenyl)-1,3-dihydro-2H-imidazol-2-ylidene) (synthesized according to Ref. (*70*), with a molar mass of 640.28 g/mol), and 18 mM DMSO (Thermo Fisher Scientific, article no. 348445000), all dissolved in 7 mL of methanol (Carl Roth, Rotisolv ≥ 99.9%, article no. T909.1).

**Acetonitrile sample**: The sample consisted of 54 mM [$^{15}$N]acetonitrile (Sigma-Aldrich, article no. 487864) and 0.33 mM [Ir(COD)(IMes)Cl] precatalyst, also dissolved in 7 mL of methanol.

**Fluoropyridine sample**: The sample consisted of 23 mM [3-$^{19}$F]pyridine (Thermo Fisher Scientific, article no. 188300050) with 1.1 mM [Ir(COD)(IMes)Cl] precatalyst, dissolved in 7 mL of methanol.

### SABRE-SHEATH

Prior to and during NMR measurements, pH$_2$ was bubbled into the 5°C cold reaction chamber at ambient pressure, with a pH$_2$ flow rate of 2 SL/h for the [1-$^{13}$C]pyruvate and [3-$^{19}$F]pyridine samples. For the [$^{15}$N]acetonitrile ZULF-COSY measurement, bubbling and measurement were performed at 2.5°C, with a pH$_2$ flow rate of 14 SL/h and a pressure of 8 bar.

During the course of the experiments, the sample concentrations gradually changed due to the evaporation of methanol.

### Apodization

As discussed in **SI chapter 3**, the sine-bell apodization function in the form:

$$\bar{a}(t) = \frac{a(t)}{\max(a(t))}, \tag{1}$$

with $a(t) = \sin\left(\frac{t}{T}\pi\right)\exp\left(-\frac{t}{T/k}\right)$, $k = \frac{T}{T_2}$, where $T$ is total measurement time, proved to be an excellent choice for improving the quality of ZULF COSY spectra. Applying such a function increased SNR by a factor of 2 and reduced ringing artifacts (see effect of apodization function on ZULF COSY in the **SI chapter 5**). The parameter $k = 4$, 6, and 4 for [1-$^{13}$C]pyruvate, [3-$^{19}$F]pyridine, and [$^{15}$N]acetonitrile was used, respectively.

### Simulations of COSY spectra

For the simulation, the moin-spin-library(*68*) (version used is available in **SI**) was utilized to generate the theoretical spectrum. The resolution, $B_0$, $B_{evo}$, bandwidth and population of spin operators were selected to match those of the experimental setup.

The simulations included the first 90° pulse, the spin evolution between the 90° pulses, the second 90° pulse, and the resulting FID, but not the hyperpolarization phase before and during the COSY experiment. As SABRE-SHEATH also hyperpolarizes multispin states, we also had to consider those for the initial state before the first 90° pulse. So, the initial state of the A₃X system, which were [1-$^{13}$C]pyruvate (A=$^1$H, X=$^{13}$C) and [$^{15}$N]acetonitrile (A=$^1$H, X=$^{15}$N) was given by the following superposition of operators:

$$\hat{\rho} = p_X \hat{I}_z^X + p_A(\hat{I}_z^{A_1} + \hat{I}_z^{A_2} + \hat{I}_z^{A_3}) + p_{2z}\hat{I}_z^X(\hat{I}_z^{A_1} + \hat{I}_z^{A_2} + \hat{I}_z^{A_3}) + p_{3z}\hat{I}_z^X(\hat{I}_z^{A_1}\hat{I}_z^{A_2} + \hat{I}_z^{A_2}\hat{I}_z^{A_3} + \hat{I}_z^{A_1}\hat{I}_z^{A_3}). \tag{2}$$

Coefficients $p$ correlate with the polarization values.(*71*)

Furthermore, we assumed imperfect 90° pulses and chose flip angles such that the simulation data aligned with the experimental data. This compensated for the spin evolution and hyperpolarization build-up at $B_{evo}$ in the experiment. The polarizations initial values and other parameters are detailed in **Table S3**. When the FIDs were obtained, the simulated signal was processed in the same way as experimental spectra, including apodization.

# Supporting materials:

Additional experimental details

Apodization

Results for [$^{15}$N]acetonitrile and [3-$^{19}$F]pyridine

Figures S1–S10

Tables S1–S3

Simulation code

## Acknowledgements:

KB acknowledges funding from the DFG (BU 2694/6-1, BU 2694/9-1). ANP and JBH acknowledge funding from the German Federal Ministry of Education and Research (BMBF) within the framework of the e:Med research and funding concept (01ZX1915C), DFG (PR 1868/3-1, PR 1868/5-1, HO-4602/2-2, HO-4602/3, GRK2154-2019, EXC2167, FOR5042, TRR287). MOIN CC was founded by a grant from the European Regional Development Fund (ERDF) and the Zukunftsprogramm Wirtschaft of Schleswig-Holstein (Project no. 122-09-053). MP acknowledges funding from the DFG (PL 576/6-1) for the Ir complex synthesis.

## Author contributions:

Conceptualization: KB, TT, ANP

Funding acquisition: ANP, KB, KS

Investigation: KB, NK, FB, ,JS, RN

Project administration: NK, KB, ANP

Resources: MP, JE, PP

Software: NK, KB, ANP, RN

Supervision: KS, KB, ANP, JBH

Visualization: NK, AO, TT, KB, RN, ANP

Writing – original draft: KB, RN, TT, ANP

Writing – review & editing: all authors

## Competing interests:

TT is a co-founders and equity holder of Vizma Life Sciences (VLS). The terms of TT's arrangement have been reviewed and approved by NC State University in accordance with its policy on objectivity in research.

# Supporting Information: Indirect Zero Field NMR Spectroscopy

Kai Buckenmaier[1,*], Richard Neumann[1], Friedemann Bullinger[1], Nicolas Kempf[1], Pavel Povolni[1], Jörn Engelmann[1], Judith Samlow[1], Jan-Bernd Hövener[2], Klaus Scheffler[1,3], Adam Ortmeier[4], Markus Plaumann[5], Rainer Körber[6], Thomas Theis[4] and Andrey N. Pravdivtsev[2]

*Corresponding author, Kai Buckenmaier kai.buckenmaier@tuebingen.mpg.de

[1] High-Field Magnetic Resonance Center, Max Planck Institute for Biological Cybernetics; Tübingen, 72076, Germany
[2] Department Section Biomedical Imaging, Molecular Imaging North Competence Center (MOIN CC), Department of Radiology and Neuroradiology, University Medical Center Kiel, Kiel University, Am Botanischen Garten 14, 24118, Kiel, Germany
[3] Departement of Biomedical Magnetic Resonance, Eberhard-Karls University; Tuebingen, 72076, Germany
[4] Departement of Chemistry and Physics, NC State University; Raleigh, 27695, USA
[5] Institute for Molecular Biology and Medicinal Chemistry, Medical Faculty, Otto-von-Guericke-University; Magdeburg, 39120, Germany
[6] Physikalisch-Technische Bundesanstalt; Berlin, 10587, Germany

## Contents

# 1. How we determined the level anticrossing field $B_{LAC}$

To achieve maximum enhancement under SABRE-SHEATH conditions, we used a simple FID readout sequence in which the hyperpolarization field, $B_{hyp}$, was systematically varied (see **Figure S1** and **Table S1**). As in the sequences described in the main manuscript, parahydrogen was continuously bubbled through the sample reactor. The acquired data was Fourier transformed and the spectra showing the absolute value of the X-nucleus signal were plotted as a function of $B_{hyp}$ (**Figure S2** upper row). To determine the optimal hyperpolarization field, $B_{LAC}$, the absolute value of the X-nucleus MR signal was integrated, and the $B_{hyp}$ that produced the highest signal was identified as $B_{LAC}$ (**Figure S2** middle row, dashed red line).

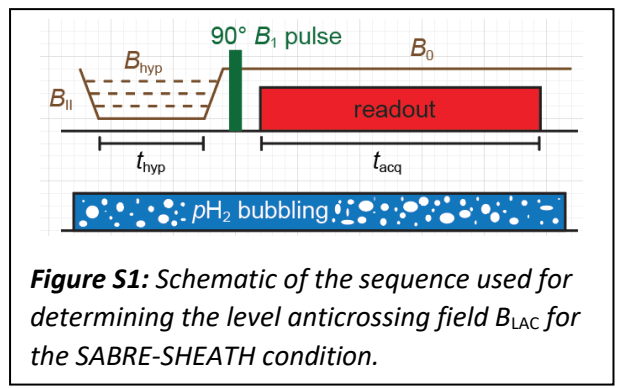

*Figure S1: Schematic of the sequence used for determining the level anticrossing field $B_{LAC}$ for the SABRE-SHEATH condition.*

**Figure S2** bottom row also shows spectra of the $^1H$ signal. For [1-$^{13}C$]pyruvate, the $^1H$ enhancement correlates directly with the [1-$^{13}C$] enhancement. In contrast, for [$^{15}N$]acetonitrile and [3-$^{19}F$]pyridine, only the $^1H$ signal of orthohydrogen (with peaks at 2200.5 Hz and 2202.0 Hz for [$^{15}N$]acetonitrile and [3-$^{19}F$]pyridine, respectively) scales with the X-nucleus signal. The proton signal of the substrate shows a different trend and signal enhancement even at $B_{hyp} = 0$.

*Table S1: Sequence parameters used for determining $B_{LAC}$.*

| substrate | $t_{hyp}$ [s] | $t_{acq}$ [s] | $B_0$ [µT] |
|---|---|---|---|
| [1–$^{13}C$]pyruvate | 4 | 2 | 6.8 |
| [$^{15}N$]acetonitrile | 4 | 4 | 51.7 |
| [3-$^{19}F$]pyridine | 4 | 4 | 51.7 |

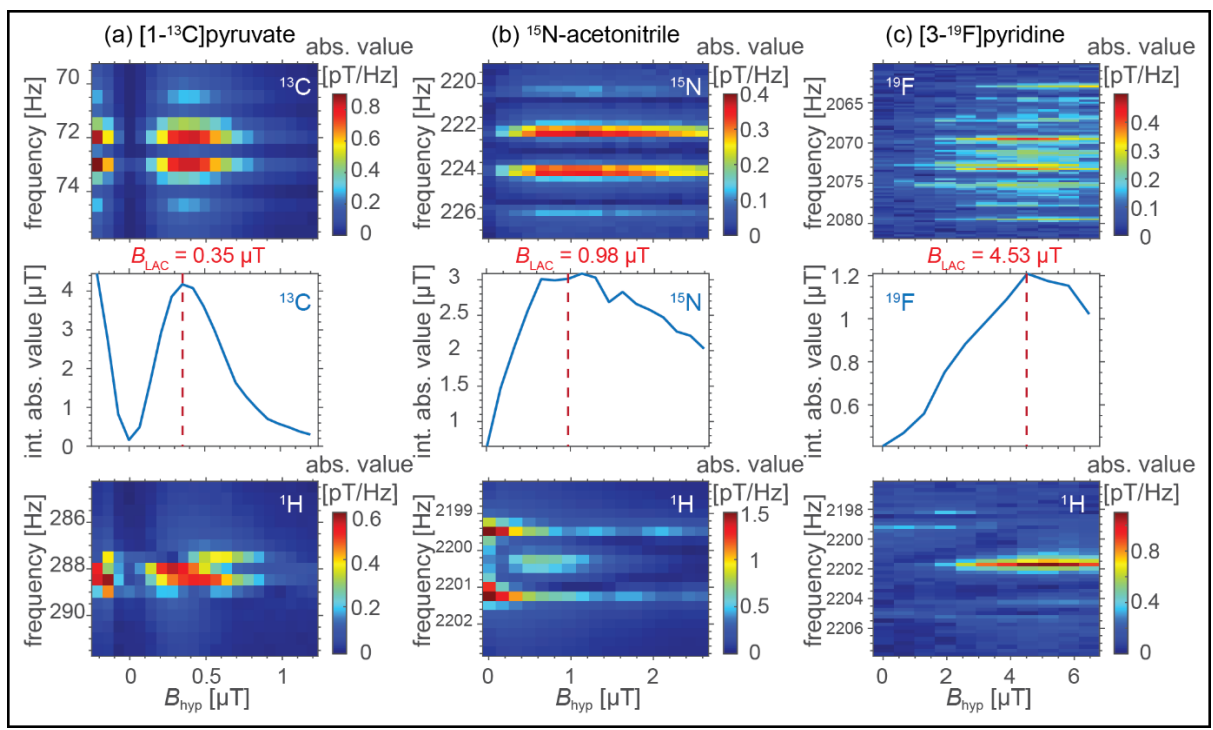

**Figure S2:** X-nucleus spectra (upper row), integrated X-nucleus MR signal (middle row) and $^1$H spectra (bottom row) of [1-$^{13}$C]pyruvate (**a**), [$^{15}$N]acetonitrile (**b**) and [3-$^{19}$F]pyridine (**c**).

## 2. Sequence parameters

The sequence parameters of all used sequences within the main manuscript are shown in the **Table S2**.

*Table S2: Sequence parameters. The table shows the sequence parameters of the discussed results.*

| Figure | $t_{hyp}$ [s] | $B_{hyp}$ [μT] | $t_{evo}$ [s] | $B_{evo}$ [nT] | $t_{acq}$ [s] | $B_0$ [μT] | $B_{ZULF}$ [μT] | $t_{total}$ |
|---|---|---|---|---|---|---|---|---|
| **ZULF FID** | | | | | | | | |
| 1a | 10 | 0.35 | – | – | 16 | – | 6.8 | 1min 45s* |
| 1b | 10 | 0.35 | – | – | 16 | – | 0 | 1min 45s* |
| **two field ZULF COSY** | | | | | | | | |
| 2 | 8 | 0.35 | 0–15 | 0 | 16 | 6.8 | – | 1h 2min |
| 3 | 8 | 0.35 | 0–9.4 | 25 | 16 | 6.8 | – | 1h 1min |
| 4 | 8 | 0.35 | 0–2.8 | 493 | 16 | 6.8 | – | 4h 44min |
| S6 | 16 | 0.982 | 0–8 | 0 | 8 | 51.7 | – | 1h 16min |
| S7 | 16 | 4.857 | 0–6 | 1305 | 8 | 51.7 | – | 3h 55min |
| S8 | 20 | 4.531 | 0–5 | 0 | 8 | 51.7 | – | 3h 58min |

* 4 averages

## 3. Apodization

Apodization is an NMR spectroscopy technique that improves the quality of spectral data by reducing artifacts and increasing the signal-to-noise ratio.(*1*) Apodization refers to the process of applying a mathematical function known as an apodization function,(*2*) to the time-domain signal before its Fourier transformation. For example, raw data have inherent noise and imperfections due to finite acquisition time, and sampling can lead to broadening spectral lines, reducing the signal-to-noise ratio (SNR). Apodization addresses these issues, resulting in better-defined peaks and more accurate quantification of spectral features.

For example, NMR signal is measured only during a narrow window that can be approximated as a rectangular function defined as:

$$f(t) = f(t) = \begin{cases} 1, & |t| \leq 0.5 \\ 0, & else \end{cases} \tag{S1}$$

The Fourier transform of a rectangle function yields:

$$F(\omega) = \int_{-\frac{1}{2}}^{\frac{1}{2}} \exp(-i\omega t)\, dt = \text{sinc}(\omega/2). \tag{S2}$$

Consequently, the Fourier transform of a finite data set can introduce artefacts - peaks that do not exist in the ideal spectrum. The data can be preprocessed by applying apodization before the Fourier transform to reduce these windowing effects. Various apodization functions are already discussed in ref. (*2*), p. 408, including a sine-bell function:

$$s(t) = \sin\left(\frac{t}{T}\pi\right), \tag{S3}$$

where $T$ is the total measurement time.

Another key reason for applying apodization is to enhance the resolution of multidimensional spectra.(*1*) The envelope of the FID typically follows an exponential decay function:

$$g(t) = M_0 \exp(-t/T_2), \tag{S4}$$

where $M_0$ is the initial amplitude and $T_2$ the transverse relaxation time.

The SNR can be improved by multiplying the FID spectrum with an exponential decay as an apodization function

$$h(t) = \exp\left(-\frac{t}{T/k}\right) \tag{S5}$$

with

$$k = \frac{T}{T_2} \tag{S6}$$

being the exponential weighting constant. For different $T_2$ a different $k$ must be chosen such that the apodization function matches the envelope of the FID spectrum.

Combining the two parts of the apodization, the final apodization function is

$$a(t) = \sin\left(\frac{t}{T}\pi\right)\exp\left(-\frac{t}{T/k}\right). \tag{S7}$$

To have comparable results the apodization function is normalized.

$$\bar{a}(t) = \frac{a(t)}{\max(a(t))}, \tag{S8}$$

**Figure S3** displays this apodization function for different weighting parameters $k$. It combines the advantages of suppressing window effects with SNR improvement.

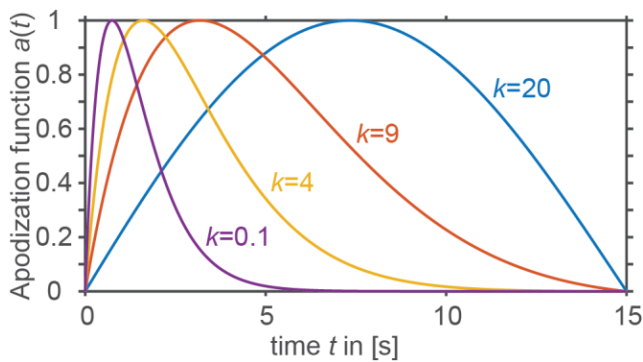

**Figure S3: Apodization functions for different weighting factors.** The figure shows the effect of the exponential weighting factor $k$ on the apodization function.

The Fourier transformation of the product of the apodization function $\bar{a}(t)$ with the FID $d(t)$, yields the spectrum with applied apodization:

$$S(\omega) = \int d(t)\bar{a}(t)\exp(-i\omega t)\,dt \tag{S9}$$

Applying the apodization function (**Figure S4** middle) to the [1-$^{13}$C]pyruvate data (**Figure S4** upper row) acquired with the ZULF COSY sequence (**Figure 1f**) results in the 2D dataset with significantly reduced noise (**Figure S4** bottom). As expected, the signal at short evolution times is attenuated by the sine bell, while the signal at long evolution times is suppressed by the exponential decay. However, a reduction in amplitude of approximately a factor of $\frac{1}{2}$ is observed, despite the fact that the SNR might be increased. Therefore, the amplitude of the apodized spectra can only be given in arbitrary units (a.u.).

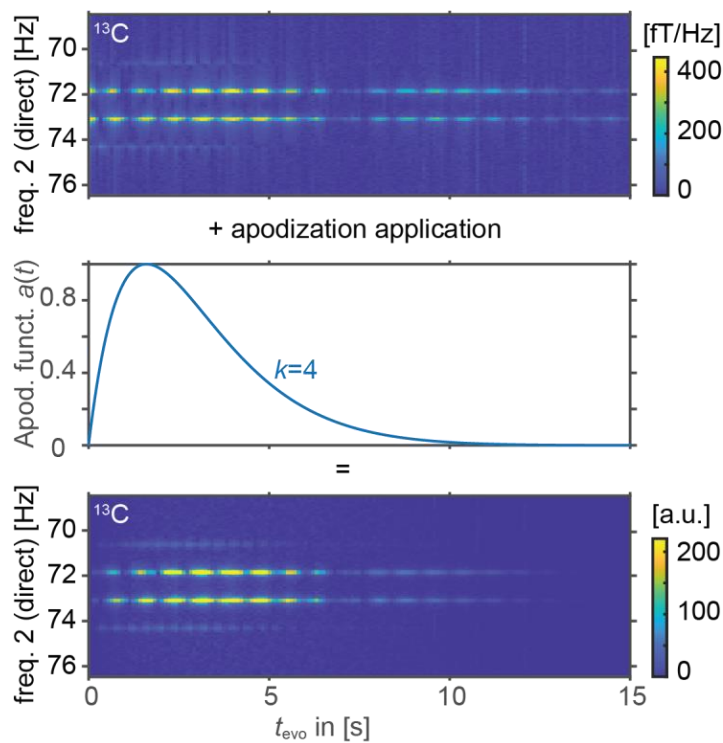

**Figure S4: Time-frequency spectrum of [1-$^{13}$C]pyruvate**. The upper figure displays the COSY spectrum with Fourier transformation applied only across the direct frequency dimension. The middle panel shows the apodization function. The bottom panel shows the product of the apodization function and the top spectrum.

Applying the Fourier transform in the indirect dimension produces the ZULF COSY spectra shown in **Figure S5**. A comparison between the ZULF COSY spectrum without apodization (**Figure S5a**) and the spectrum with apodization applied (**Figure S5b**), reveals significant improvements.

The application of apodization results in enhanced SNR by reducing the noise by an estimated factor of 2, and diminished ringing artifacts. Notably, in the projection along the indirect frequency dimension, the outer peaks of the quintuplet become clearly visible in the apodized spectrum. However, due to the modification of the measurement data by apodization, the amplitude is altered, and therefore given in a.u. (**Figure S5**).

In ZULF COSY figures of the results section within the main manuscript and all subsequent ZULF COSY figures, apodization was applied, allowing for a more refined and detailed analysis. For [1-$^{13}$C]pyruvate, $k = 4$ ([3-$^{19}$F]pyridine, $k = 6$ and for [$^{15}$N]acetonitrile, $k = 4$) proofed to be an optimized choice.

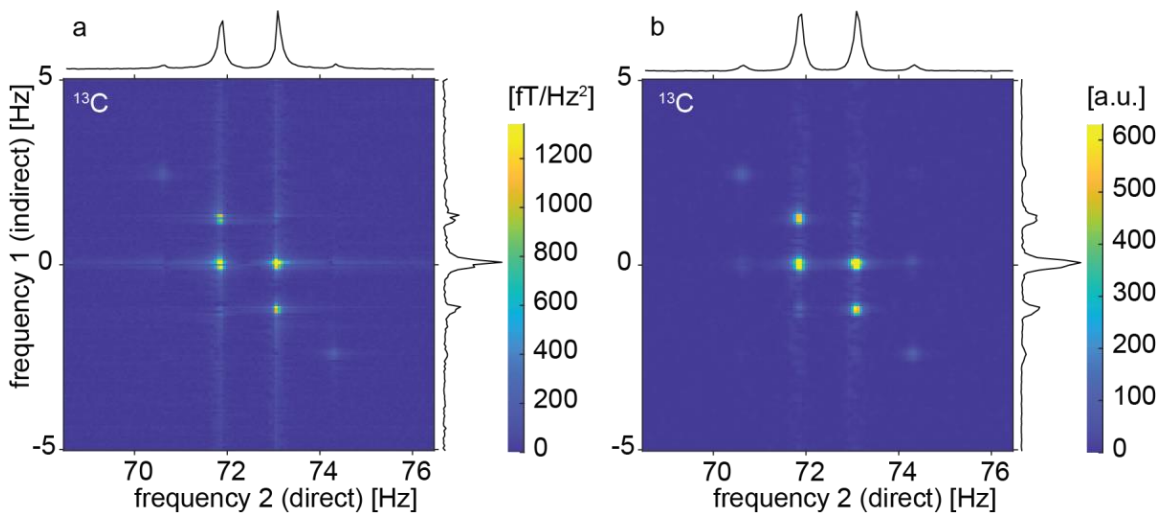

**Figure S5: Effect of apodization to the ZULF COSY spectrum of [1-$^{13}$C]pyruvate.** The figure illustrates the effect of apodization on two-dimensional spectra. It is evident that the spectrum without apodization (**a**) does not display the outer peaks in the indirect dimension projection of the $^{13}$C whereas these peaks become clearly observable in the apodized spectrum (**b**). The SNR enhancement resulting from the application of apodization is apparent in both the two-dimensional spectrum and its projections.

# 4. ZULF COSY spectra in the ZULF and Zeeman regime of [$^{15}$N]acetonitrile and [3-$^{19}$F]pyridine

Additionally, to the [1-$^{13}$C]pyruvate, spectra of [$^{15}$N]acetonitrile and [3-$^{19}$F]pyridine will be shown. While [$^{15}$N]acetonitrile is an $A^3X$ system similar to [1-$^{13}$C]pyruvate and thus the results show the same pattern, [3-$^{19}$F]pyridine is a far more complex molecule and so is the obtained spectrum.

## 4.1. ZULF regime ZULF COSY spectrum of [$^{15}$N]acetonitrile

The results of ZULF COSY [$^{15}$N]acetonitrile are as anticipated from the [1-$^{13}$C]pyruvate data as both, [$^{15}$N]acetonitrile and [1-$^{13}$C]pyruvate, are $A^3X$ systems.

The ZULF evolution field of $B_{evo} = 0\,\text{nT}$ results in a zero-field spectrum, in the indirect dimension, where both the $^{15}$N and $^{1}$H projections display a well-defined quintuplet with peaks at 0, $\pm J$, and $\pm 2J$ frequencies. Note that for [$^{15}$N]acetonitrile $^3J_{NH}$ is about 1.75 Hz.

For this experiment again, it was necessary to use further shimming coils. The shimming process itself was performed by acquiring ZULF FID spectra while varying the shimming fields accordingly.

In the direct dimension projection, the quadruplet peaks of $^{15}$N, separated by the $J$-coupling frequency, as well as the doublet of $^{1}$H, also split by the $J$-coupling frequency, can be observed.

To align the simulation with experimental results, exactly the same simulation parameter as for the [1-$^{13}$C]-pyruvate are selected. The parameters can be found in **Table S3**. The peak positions in both the measured and simulated data show excellent agreement (**Figure S6**).

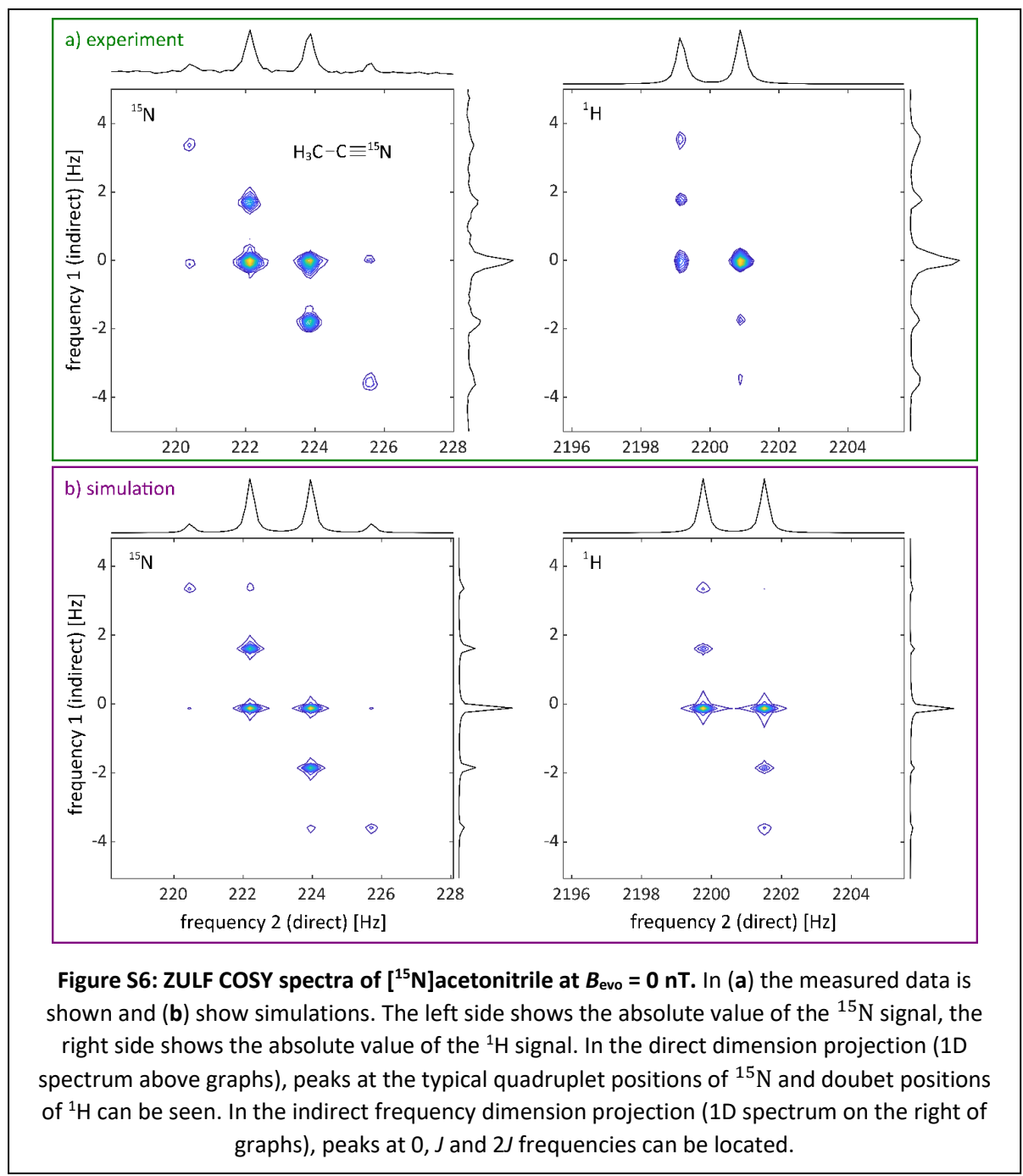

**Figure S6: ZULF COSY spectra of [$^{15}$N]acetonitrile at $B_{evo}$ = 0 nT.** In (**a**) the measured data is shown and (**b**) show simulations. The left side shows the absolute value of the $^{15}$N signal, the right side shows the absolute value of the $^1$H signal. In the direct dimension projection (1D spectrum above graphs), peaks at the typical quadruplet positions of $^{15}$N and doubet positions of $^1$H can be seen. In the indirect frequency dimension projection (1D spectrum on the right of graphs), peaks at 0, *J* and 2*J* frequencies can be located.

### 4.2. Zeeman regime ZULF COSY spectrum of [$^{15}$N]acetonitrile

The magnetic field strength of the evolution field $B_{evo} = 1.3$ µT falls within a range where the ZULF COSY spectrum displays Zeeman dominated interaction. In this regime, the 2D spectrum exhibits symmetrical patterns, and the peaks can be attributed to direct peaks and cross-peaks of $^{15}$N. However, the direct and cross-peaks within the indirect direction are located at $^{15}$N Larmor frequencies of $B_{ZULF}$ (5.7 Hz respectively, **Figure S7**, brown boxes). The proton peaks are only observable as direct peaks at the proton Larmor frequencies of $B_0$ and the proton Larmor frequencies of $B_{evo}$ (15.6 Hz). The cross-peaks of the proton are not observable. The areas where the proton peaks can be expected are marked by the blue boxes in **Figure S7**.

In the direct dimension, the spectrum reveals the anticipated quadruplet peaks for $^{15}N$, which are separated by the J-coupling frequency $3J_{NH}$ = 1.75 Hz. Similarly, for $^1H$ a doublet structure is observed, also spaced by the J-coupling frequency.

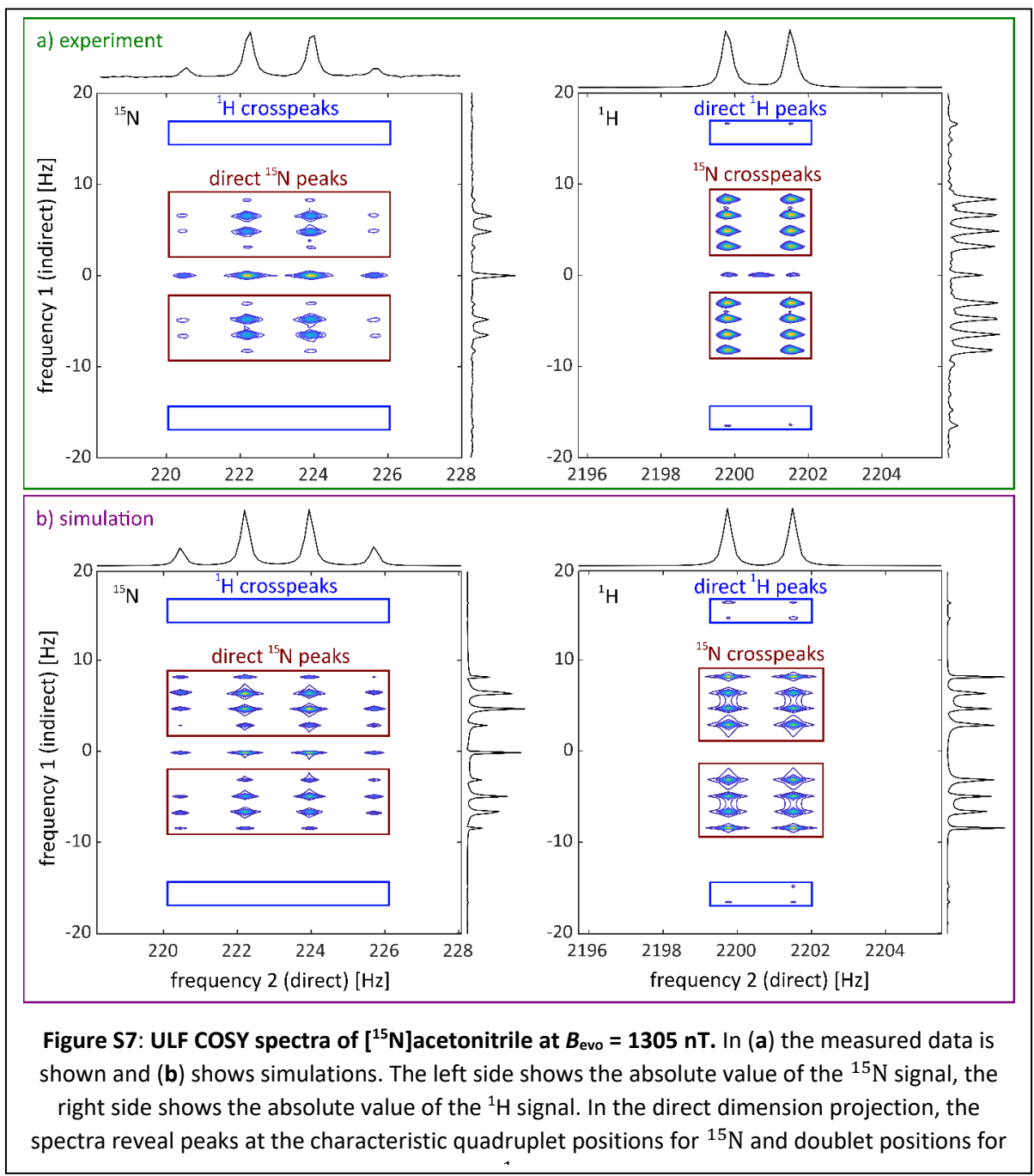

**Figure S7**: **ULF COSY spectra of [$^{15}N$]acetonitrile at $B_{evo}$ = 1305 nT.** In (**a**) the measured data is shown and (**b**) shows simulations. The left side shows the absolute value of the $^{15}N$ signal, the right side shows the absolute value of the $^1H$ signal. In the direct dimension projection, the spectra reveal peaks at the characteristic quadruplet positions for $^{15}N$ and doublet positions for

Similarly, as reported for the [1-$^{13}C$]pyruvate, there are high peaks at 0 Hz in the indirect dimension of the experimental $^{15}N$ signal (**Figure S7a**). During the evolution time between the two 90° pulses, $B_{ZULF}$ is set to 1.3 µT, which is close to the $B_{LAC}$ field of 1 µT. As a result, longitudinal polarization builds up during this period resulting in peaks located at 0 Hz of the indirect direction.

To align the simulation with experimental results, a proton polarization.$p_A = 0$, and $^{15}$N polarization $p_X = 1$ were selected.

### 4.3. ZULF regime ZULF COSY spectrum of [3-$^{19}$F]pyridine

Using [3-$^{19}$F]pyridine as ligand in the ZULF COSY experiment yields a complex spectrum (**Figure S8**). It is observable that the spectrum has similar characteristics as the [$^{15}$N]acetonitrile and [1-$^{13}$C]pyruvate spectra, e.g., the asymmetric pattern and the indirect dimension projection being placed around 0 Hz. But the spectrum is far too complex to make further assumptions, due to the complex $J_{FH}$ coupling pattern (2 different $^3J_{FH}$, as well as $^4J_{FH}$ and $^5J_{FH}$).(*3*)

Again, there is a high orthohydrogen peak at the symmetry center of the pattern in the experimental data of the $^1$H data (**Figure S8a**, red box).

To align the simulation with experimental results, a proton polarization $p_A = 0.29$, and $^{19}$F polarization $p_X = 0.71$ were selected. With these settings simulation data were obtained that almost perfectly aligns in the direct and indirect dimension projections with the experimental data.

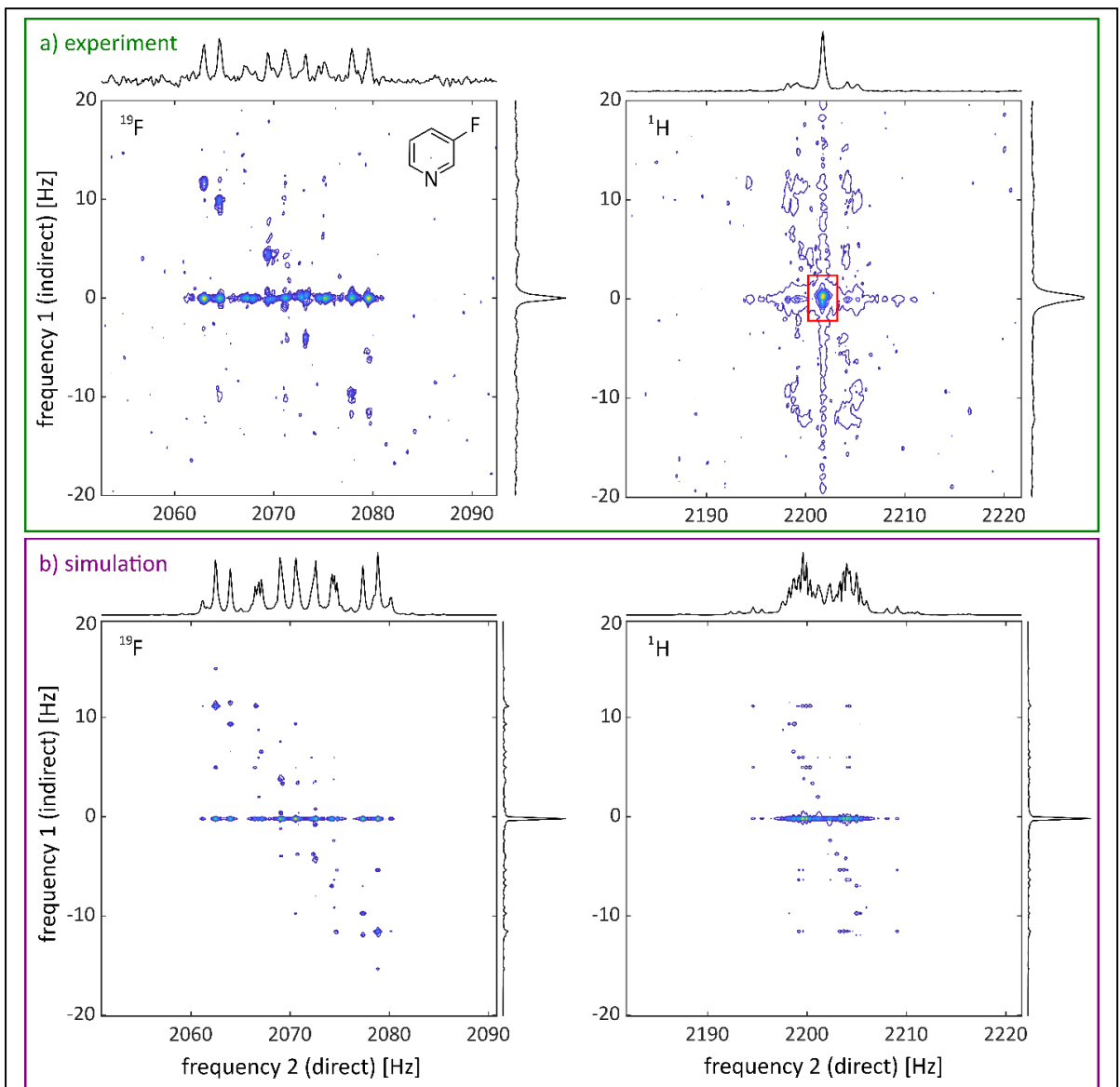

**Figure S8 : ZULF COSY spectra of [3-$^{19}$F]-pyridine at $B_{evo}$ = 0 nT.** In (**a**) the measured data is shown and (**b**) show simulations. The left side shows the absolute value of the $^{19}$F signal, the right side shows the absolute value of the $^{1}$H signal. In the direct dimension projection (1D spectrum above graphs), peaks at the typical nine positions of $^{19}$F and doublet positions of $^{1}$H can be observed.

# 5. Effect of apodization on [$^{15}$N]acetonitrile and [3-$^{19}$F]pyridine ZULF COSY spectra

In order to show the robustness of the apodization formalism and function, the effect on the unfiltered data of [$^{15}$N]acetonitrile and [3-$^{19}$F]pyridine are presented below. Due to the fact that the signal of [$^{15}$N]acetonitrile (**Figure S9**) is much lower than the signal of [1-$^{13}$C]pyruvate the effect of the apodization is even more observable. Applying the apodization on the [3-$^{19}$F]pyridine data (**Figure S10**), where the SNR is even lower than for [$^{15}$N]acetonitrile leads to line broadening and a reduction of resolution due to the high exponential weighting factor $k = 6$. This effect can also be observed for the [$^{15}$N]acetonitrile. The positive effect of the apodization is the improvement of the SNR and reduction of ringing, observable in both figures.

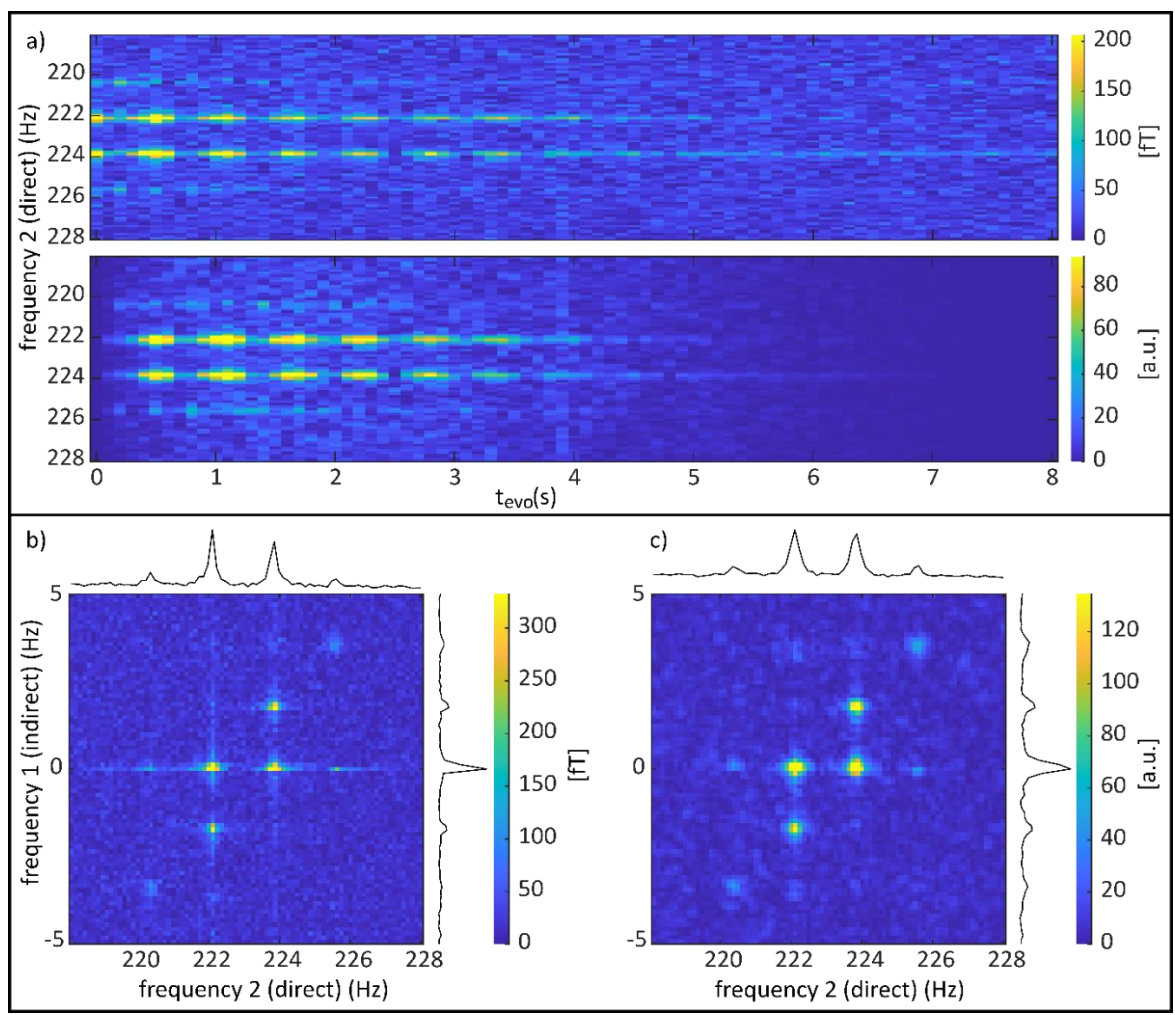

**Figure S9: Effect of the apodization on [$^{15}$N]acetonitrile data.** The figure shows the difference between the spectra with and without apodization being applied. In (**a**) the direct dimension frequency dependent on the evolution time $t_{evo}$ is shown. The upper figure shows the spectrum without apodization applied while the bottom figure shows the apodized spectrum. In (**b**) the ZULF COSY spectrum in the ZULF regime and in (**c**) the spectrum with applied apodization are displayed.

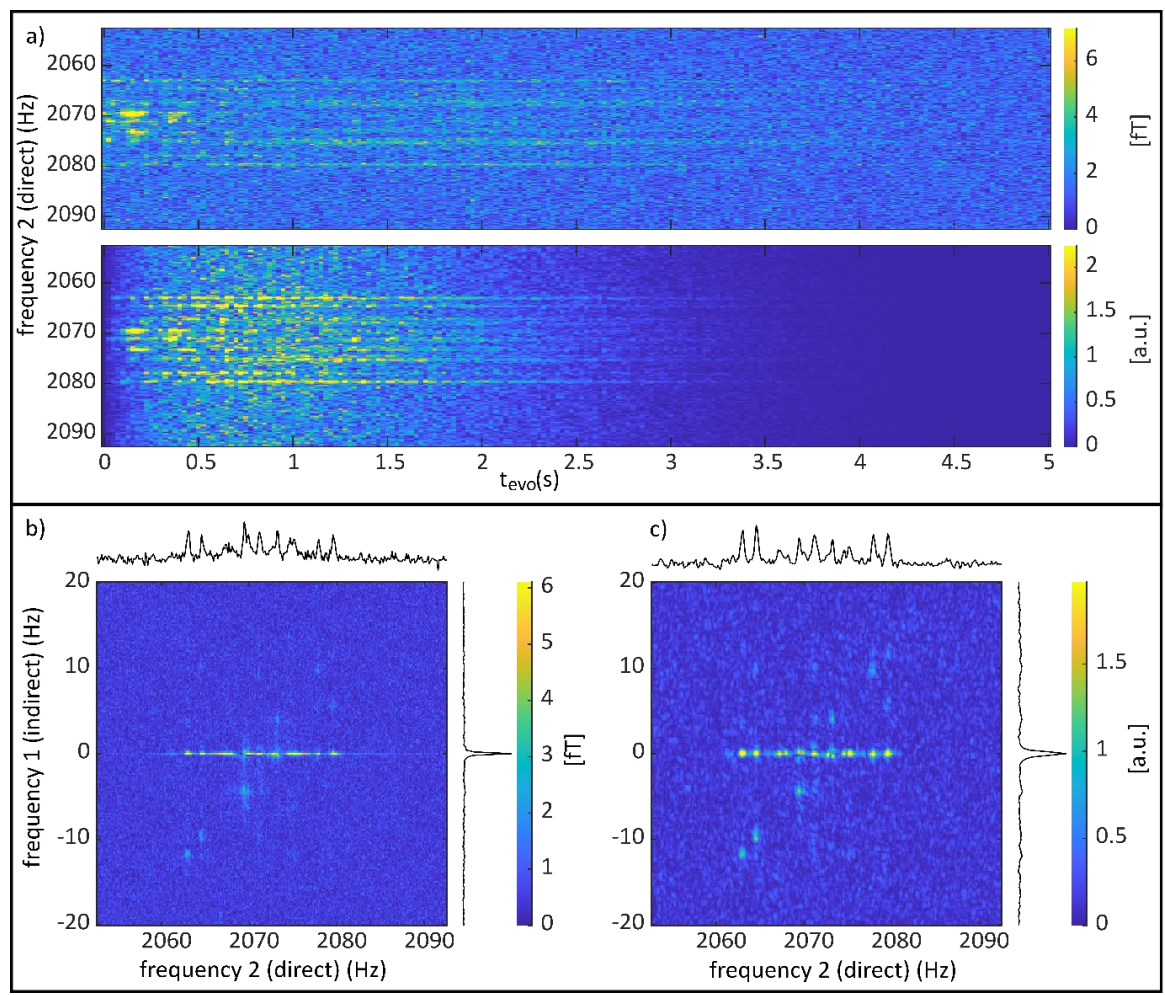

**Figure S10: Effect of the apodization on [3-$^{19}$F]-pyridine data.** The figure shows the difference between the spectra with and without apodization being applied. In (**a**) the direct dimension frequency dependent on the evolution time $t_{evo}$ is presented. The upper figure shows the spectrum without apodization applied while the bottom figure shows the apodized spectrum. In (**b**) the ZULF COSY spectrum in the ZULF regime and in (**c**) a spectrum measured under the same condition but with applied apodization is presented.

## 6. Simulation parameters

Quantum coherences up to the third order for [1-$^{13}$C]pyruvate and [$^{15}$N]acetonitrile were populated to obtain the optimal results in the simulation. For [3-$^{19}$F]pyridine only the first order quantum coherences were populated. The flip angle also had to be adjusted. The used simulation parameters are shown in the following **Table S3**.

*Table S3: Simulation parameters. The table shows the simulation parameters of the discussed results.*

| Figure | Flip angle [°] | $p_A$ | $p_X$ | $p_{2z}$ | $p_{3z}$ | $p_{4z}$ |
|---|---|---|---|---|---|---|
| 2 | 60 | 0.09 | 0.23 | 0 | -0.68 | 0 |
| 3 | 100 | 0.074 | 0.926 | 0 | 0 | 0 |
| **4** | 100 | 0 | 1 | 0 | 0 | 0 |
| **S6** | 60 | 0.09 | 0.23 | 0 | -0.68 | 0 |
| **S7** | 100 | 0 | 1 | 0 | 0 | 0 |
| **S8** | 60 | 0.29 | 0.71 | 0 | 0 | 0 |